\documentclass[12pt]{article}\usepackage[]{epsfig}
\begin{document}
{\LARGE \begin{center}
The CRESST Dark Matter Search 
\end{center}}
\noindent

M.~Bravin$^1$,
M.~Bruckmayer$^3$,
C.~Bucci$^4$,
S.~Cooper$^{3}$,
S.~Giordano$^1$,
F.~v.Feilitzsch$^2$,
J.~H\"ohne$^2$,
J.~Jochum$^2$,
V.~J\"orgens$^4$,
R.~Keeling$^3$,
H.~Kraus$^3$,
M.~Loidl$^1$,
J.~Lush$^1$,
J.~Macallister$^3$,
J.~Marchese$^3$,
O.~Meier$^1$,
P.~Meunier$^1$,
U.~Nagel$^2$
T.~N\"ussle$^2$,
F.~Pr\"obst$^1$,
Y.~Ramachers$^3$,
M.~Sarsa$^2$,
J.~Schnagl$^2$,
W.~Seidel$^1$
I.~Sergeyev$^{1,5}$,
M.~Sisti$^1$,
L.~Stodolsky$^1$,
S.~Uchaikin$^{1,5}$,
L.~Zerle$^1$
\\[2mm]

\noindent
{\it\small $^1$ Max-Planck-Institut f\"ur Physik, F\"ohringer Ring
6,
D-80805 Munich, Germany}
\\
{\it\small $^2$ Technische Universit\"at M\"unchen, Physik
Department,
D-85747 Munich, Germany}
\\
{\it\small $^3$ Univeristy of Oxford, Physics Department, Oxford
OX1 3RH, UK}
\\
{\it\small $^4$ Laboratori Nazionali del Gran Sasso, I-67010
Assergi, Italy}
\\
{\it \small $^5$ Permanent Address: Joint Institute for Nuclear Research, Dubna, 
141980, Russia}
\\[0.5cm]

\begin{abstract}
We discuss the short and long term perspectives of the
CRESST
(Cryogenic Rare Event Search using Superconducting
Thermometers)
project and
present
the current status of the experiment and new results concerning
detector development.  In the search for elementary particle dark
matter,  CRESST is presently the  most advanced
deep underground,  low background, cryogenic facility.  The basic
technique involved is to search for WIMPS (Weakly Interacting
Massive Particles) by the
measurement  of non-thermal
phonons, as created by WIMP-induced nuclear recoils.  
Combined with our newly developed method for the 
simultaneous measurement of
scintillation light,  strong  background discrimination is
possible, resulting in a substantial increase in WIMP detection
sensitivity. 
This will allow  a test  of the
reported  positive evidence for  a WIMP signal by the DAMA
collaboration in the
near future. In the
long term, the present CRESST set-up permits the
installation of
a detector mass up to 100 kg. 
\par
In contrast
to other projects, CRESST technology allows the 
employment of a large variety of detection materials.
This offers
a powerful tool in establishing a WIMP signal and
in investigating
WIMP properties  in the event of a positive signal. 
\end{abstract}

PACS: 95.35+d, 29.40 \hfill\break
Keywords: Dark Matter, Direct Dark Matter Detection, Radiation Detector, 
Cryogenic Detector \hfill\break

\newpage


\section{CRESST and the Dark Matter Problem}

After a long period of development, cryogenic detectors are now
coming on line and in the next years  will deliver
significant results in particle-astrophysics and weak interactions.
The stable operation of a kilogram of detecting material in the
millikelvin range over long time periods by CRESST, as well
as similar work by other collaborations, has confirmed the hopes
that
large mass cryogenic detectors are feasible.
CRESST is presently the  most advanced deep underground, low
background, cryogenic facility.
Other major projects are the CDMS project in Stanford, the
EDELWEISS project at Frejus,
the Milano $\beta  \beta$ project in Gran Sasso, the ROSEBUD
experiment at Canfranc, the Tokyo Cryogenic Dark Matter Search and the
ORPHEUS project at Bern \cite{ltd7}.

The goal of the  CRESST project  is the direct detection
of elementary particle dark matter and the elucidation of its
nature.  The search for  Dark Matter and the understanding of its nature
remains one of the central and most fascinating problems of our
time in physics, astronomy and cosmology. There is strong evidence
for it on all scales, ranging from dwarf galaxies, through spiral
galaxies like our own, to large scale structures. The history of
the universe is difficult to reconstruct without it, be it big bang
nucleosynthesis \cite{nuclsyn} or the formation of structure \cite{dod}.

 The importance of the search for dark matter in the form of
elementary particles, created in the early stages of the universe,
is underlined by the recent weakening of the case for other forms
 such as MACHOS, faint stars and black holes \cite{workshops}.
 Particle physics provides a well motivated candidate  through the
assumption that the lightest supersymmetric (SUSY) particle, the
`neutralino´, is some combination of neutral particles arising in 
the theory and it is possible to find many candidates obeying
cosmological and particle physics constraints. Indeed,
 SUSY models contain  many parameters and many
assumptions, and by relaxing various simplifying  assumptions one
can find candidates  in a wide mass range \cite{pok}.
 Generically, such particles are called WIMPS (Weakly Interacting
Massive Particles), and are to be distinguished from proposals
involving very light quanta such as axions. WIMPS are expected to
interact with ordinary matter by elastic scattering on nuclei and
all direct detection schemes have focused on this possibility.

Conventional methods for  direct detection  rely on the ionisation
or scintillation
caused by the recoiling nucleus. This leads to certain limitations
connected with the relatively high energy involved in producing 
electron-ion or electron-hole pairs 
and with the sharply decreasing efficiency of ionisation
by slow nuclei.
Cryogenic detectors use the much lower energy
excitations, 
such as phonons,  and while
conventional methods are probably close to their limits, cryogenic
technology can still 
make great improvements.
Since the principal physical effect of a WIMP nuclear recoil is the
generation of phonons, cryogenic calorimeters are well suited for WIMP detection 
and, indeed, the first proposals to search for
dark matter particles were inspired by early work on cryogenic
detectors \cite{gw}.
Further, as we shall discuss below, when this technology  is
combined with   charge or light
detection the resulting background suppression leads to a powerful
technique  to search for the rare nuclear recoils due to WIMP
scatterings.

The detectors developed by the CRESST collaboration consist of a
dielectric crystal (target or absorber)
with a small  superconducting film  (thermometer)
evaporated onto the surface. When this film is held at a temperature
in the middle of its superconducting  to normal conducting phase transition, it   
functions as
a highly sensitive thermometer. The detectors presently employed in
Gran Sasso use tungsten (W) films and sapphire ($Al_2O_3$)
absorbers,
 running near 15 mK. It is important for the following, however, to
realise that the technique can also be applied to a variety of
other materials.
 The small
change in  temperature of the superconducting  film resulting from
an energy deposit in the absorber  leads to a relatively large
change in the film's resistance. This change in resistance 
is measured with a SQUID.
To a good approximation, the high frequency phonons created by an
event
do not thermalise in the crystal before being directly absorbed in
the superconducting film \cite{model}.  Thus
the energy resolution is  only moderately dependent on the size of
the crystal, and scaling up to
 large detectors of some hundred gramms or even kilogramms is feasible.
The high sensitivity of this system also allows us to use  a small
separate detector of the same type to see the light  emitted  when
the absorber is a scintillating crystal.

\section{Present Status of CRESST}

The task set for the first stage of CRESST was to show the
operation of 1\,kg of
sapphire in the  millikelvin range, with a threshold of 500\,eV
under low background conditions \cite{proposal93}.
Meeting this goal involved  two major tasks:
\begin{itemize}
\item The setting up of a low background, large volume, cryogenic
installation and
\item the development of massive, low background detectors with
low
energy thresholds.
\end{itemize}

\subsection{CRESST Installation in the Gran Sasso Laboratory (LNGS)}

The central  part of the CRESST low background facility at the LNGS
is the cryostat.
The design of this cryostat had to combine the requirements of low
temperatures with those of a low background.
The first generation cryostats in this field were conventional
dilution refrigerators where some of  the
 materials were screened for radioactivity. However,
due to cryogenic requirements some non-radiopure materials, for example 
stainless
steel,  cannot be completely avoided.
Thus  for   a second generation low background cryostat, a design
was chosen in which
a well separated `cold box´ houses the experimental volume at some
distance from the cryostat. 
The cold box is  constructed entirely of low background
materials, without any compromise. It is surrounded by  shielding
consisting of 20\,cm of lead
and 14\,cm of copper.
The cooling power of the dilution refrigerator is transferred to
the
cold box by a 1.5 meter long cold finger protected by thermal radiation
shields, all  of low background copper.
The experimental volume can house up to 100\,kg of target mass.
The cold box and  shielding are installed in a clean room area with
a measured clean room class of 100.
For servicing, the top of the cryostat can be accessed from the
first floor
outside the clean room.
This situation of a second generation cryostat in a high quality
clean room,
 deep underground in the LNGS,
presently makes this instrument  unique in the world.

The installation is now complete and entering into full
operation.
The system demonstrated  its high reliability by running for more
than a
year with a prototype cold box made of normal copper.
Runs with a new low background cold box in the fall of 1998
showed stable operation for a period of months.
At present four 262\,g detectors are in the experimental volume,
performing first measurements under low background conditions.
First results of this run have shown that at energies above 30\,keV,
the
counting rate is on the order of a few counts/ (kg\,keV \,day) and above
100\,keV
below 1\,count/(kg \,keV\,day)
There are strong indications that the low energy part of the
spectrum  was dominated
by external disturbances such as mechanical vibrations or
electromagnetic interference.
We are working to correct this in future runs.

\subsection{Detector Development}

\begin{figure}[tb]
\vspace*{-1cm}
\begin{center}
\mbox{\epsfig{file=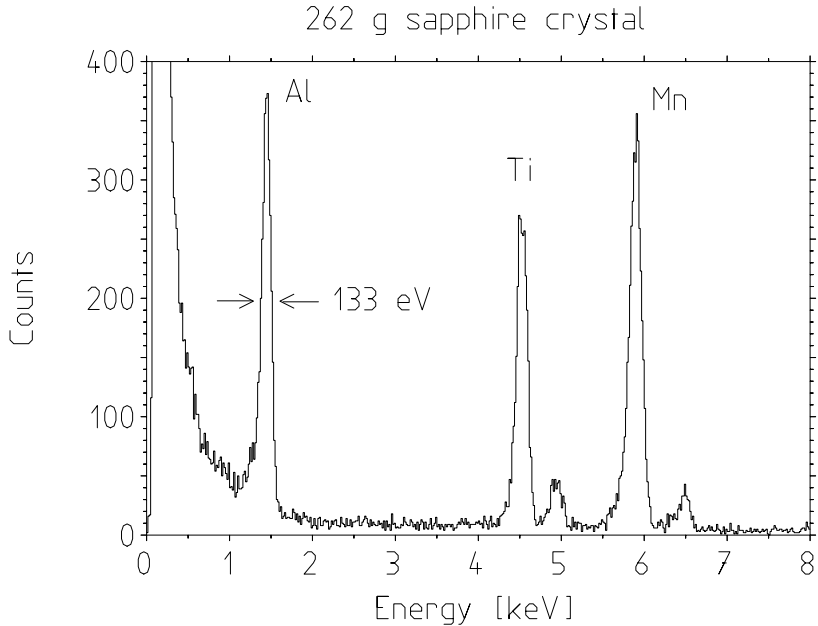,height=12cm,angle=90}}
\end{center}
\caption[]{Pulse height spectrum of a 262\,g detector
operated with thermal feedback and an X-ray fluorescence
source~\cite{colling}
installed inside the cryostat
to provide the X-ray lines of Al, Ti, and Mn.
The large background towards lower energies,
which was not present in our earlier spectra \cite{sisti},
is attributed to damage later noticed
to the thin Al sheet meant
to absorb Auger electrons from the source.}
\label{fig-133eV}
\end{figure}

The  CRESST collaboration is among the  pioneers  of
cryogenic detector development.
 Present CRESST detectors have by far the highest sensitivity per
unit
mass of any cryogenic device now in use.
Figure \ref{fig-133eV} shows the spectrum of an X-ray fluorescence
source
measured with a 262\,g CRESST sapphire detector, as presently being
used, showing an energy resolution of  133\,eV at 1.5\,keV.

These 262\,g detectors were developed  by scaling up a 32\,g
sapphire detector
\cite{colling}.
Due to optimised design, and because of the non-thermalization of
the phonons as explained in the introduction, this scaling-up could
be achieved without
loss in sensitivity.
Further developments for the next detector generation
are in progress.
\begin {itemize}
\item 
In order to improve linearity, dynamic range and
time response, a  mode of operation with thermal feedback  was
developed and  successfully operated  with the present CRESST
detectors.

\item 
For another thermometer type, the  iridium-gold proximity
sandwich, fabrication improvements now allow the application of
these thermometers with a wide choice of
absorber materials,
even for low melting point materials such as germanium.
A germanium detector with a mass of 342\,g is in preparation.

\item 
To further increase the energy sensitivity of the  detectors
 we have also  developed phonon collectors. The collectors provide
a large collection area for phonons while retaining a small
thermometer. This allows a more rapid collection of the phonons and
so an  increase in sensitivity.
 This concept can be applied to all  detector types and is
especially of interest with regard to scaling up the size of the
detectors.

\item  
Passive techniques of background reduction --  radiopure
materials and a low background environment -- are of course
imperative in work of this type. However, there is a remaining
background  dominated by
$\beta$ and $\gamma$ emissions from nearby
radioactive contaminants. These produce exclusively  electron
recoils in
the detector. In contrast WIMPs, and of course also neutrons, lead
to nuclear recoils.
Therefore, dramatic improvements in sensitivity  are to be expected
if, in addition to the usual passive shielding, the
detector itself is capable of distinguishing  electrons from
nuclear recoils and rejecting them.

\end {itemize}

\subsection{Simultaneous Phonon and Light Measurement}

We have recently developed a system, presently using CaWO$_4$
crystals as the absorber,  where a measurement of scintillation 
light is carried out in parallel to the  phonon
detection. We find that these devices clearly discriminate 
nuclear recoils from electron recoils.


\begin{figure}[htb]
\begin{center}
\epsfig{file=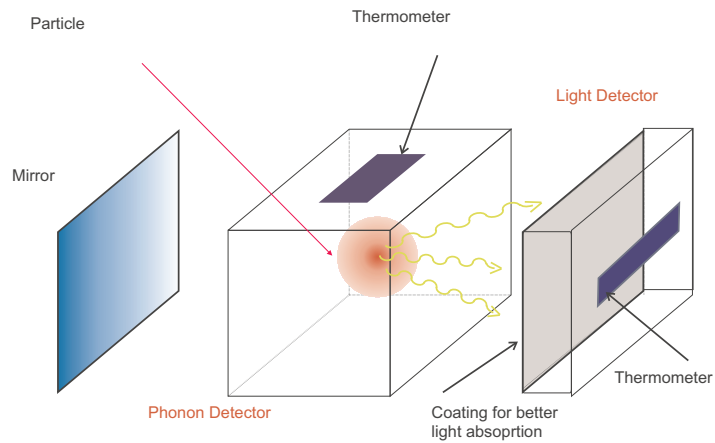,height=6cm,angle=0} 
\end{center}
\caption{Schematic view of the arrangement used for the simultaneous light and 
phonon detection}
\label{fig:schema}
\end{figure}

The system is
shown schematically in fig.~\ref{fig:schema} .  It consists of two independent 
detectors, each  of the
CRESST type:
A scintillating  absorber with a tungsten superconducting phase
transition thermometer on it,  and  a   similar but smaller detector placed next 
to it  to detect the
scintillation light from the first detector. 
A detailed description is given in \cite{cawopaper}. 
Both detectors detectors
were made by standard CRESST techniques and were operated at about 12mK. 
The CaWO$_4$ crystal 
was irradiated with photons and simultaneously with electrons. 

\begin{figure}[tb]
\hspace*{-0cm}
\mbox{\epsfig{file=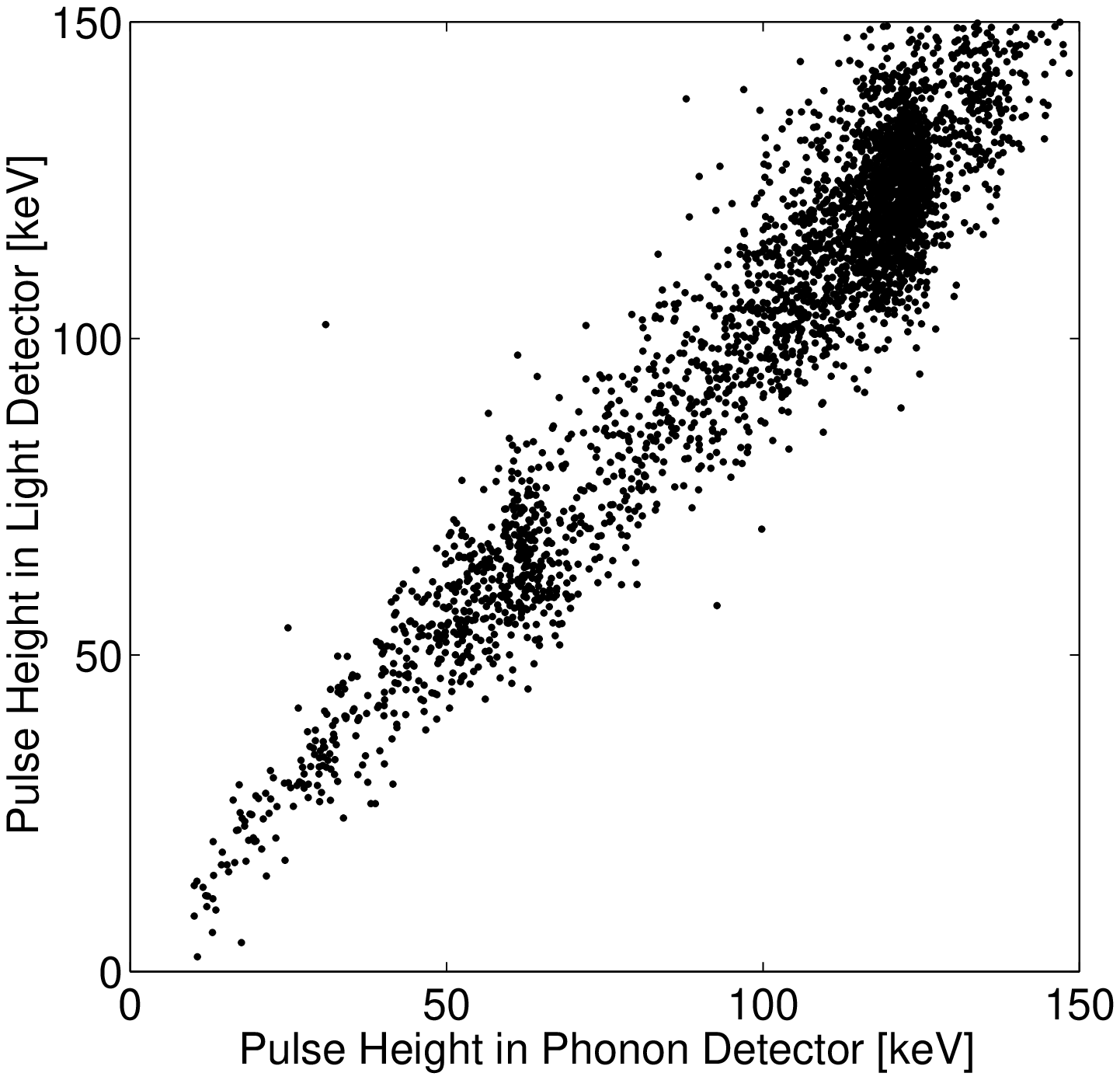,width=7.8cm,angle=0}\hspace*{-
0cm}\epsfig{file=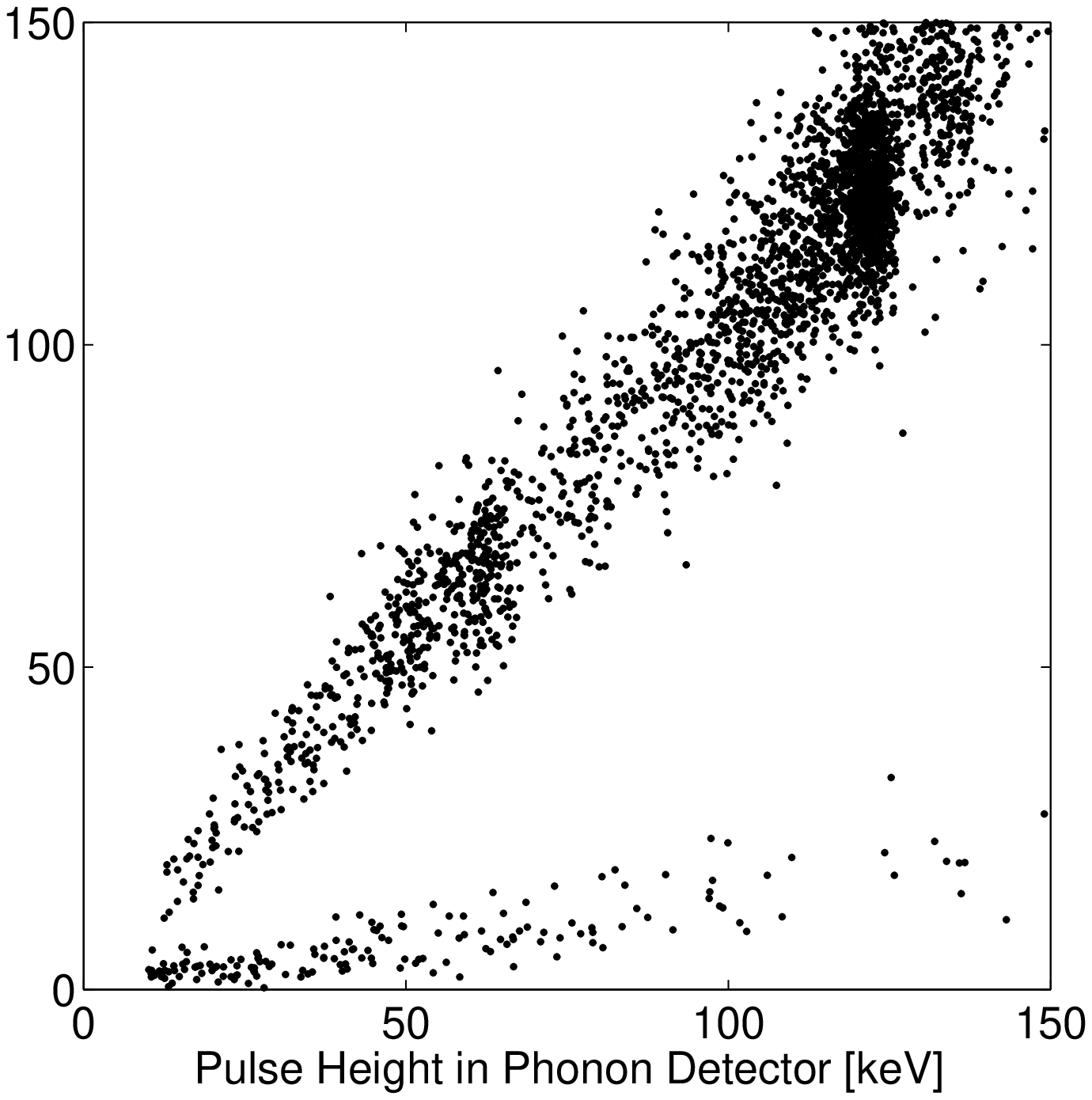, width=7.8cm,angle=0} }
\caption{Pulse height in the light detector versus pulse height in
the phonon detector. The  scatter plot on the left side has been  measured with 
an electron-  and a photon
source, while a neutron source was added on the right. }
\label{fig:gamcorr}
\end{figure}

The left plot in fig.~\ref{fig:gamcorr} shows a scatter plot of the pulse 
heights observed in the light detector versus
the pulse height observed in the phonon detector. A clear 
correlation between the light and phonon signals is observed. The right hand 
plot shows the result of an additional irradiation with neutrons from an 
Americium-Beryllium source.  
A second line can be seen  due to  neutron-induced nuclear recoils. It
is to be observed that electron
and nuclear recoils can be clearly distinguished down to a
threshold of 10keV.

The leakage of some electron recoils  into the nuclear recoil line
gives the
electron recoil rejection according to the quality factor of ref. 
~\cite{gaitskell}. 
A detailed evaluation yields a rejection factor of 98\% in the energy range 
between 10\,keV and 20\,keV, 99.7\% in the range between 15\,keV and 25\,keV and 
better than 99.9\% above 20\,keV.





The intrinsic background in our  CaWO$_4$ crystals is now being
measured in
the new Munich low background laboratory.
No contamination was found as of this writing.
The present limits
are 45 counts/(kg\,keV\,day)  for the thorium chain and 6\,counts/(kg\,keV\,day) 
for 
the uranium chain in the energy region relevant for the WIMP dark matter 
search.

\section{Next Steps for CRESST}

All our detectors, including those measuring  scintillation light,
use superconducting phase transition thermometers with SQUID
readout and can be run in the present set-up.
The CRESST cold box is
designed to house detectors of various types,
up to a total mass of about 100\,kg.

 Due to the complementary detector concepts of low threshold
calorimeters on the one hand and  detectors with the simultaneous
measurement of light and phonons on the other, CRESST can cover a
very wide range of WIMP masses.

\subsection{Low Mass WIMPs}

\begin{figure}[t]
\vspace*{-1cm}
\begin{center}
\mbox{\epsfig{file=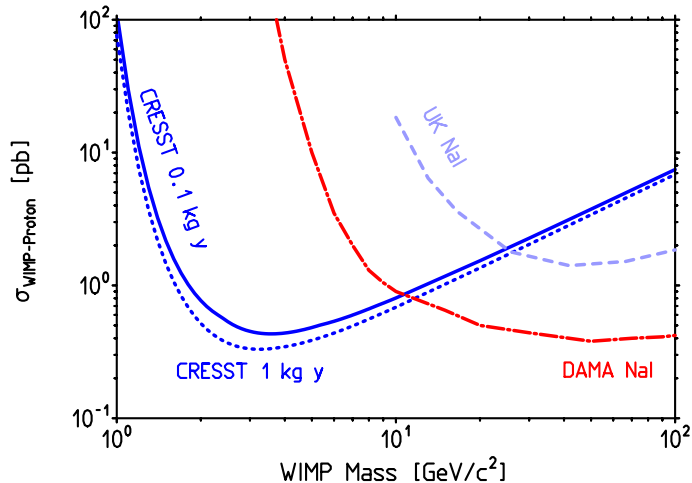,height=10cm,angle=90}}
\end{center}
\caption[]{
Equivalent WIMP-proton cross section limits (90\% CL) for spin
dependent
interactions as a function of the WIMP-mass, as expected for the
present
CRESST sapphire detectors with a total mass of 1 kg. The
expectation is based
on a threshold of 0.5 keV, a background of 1\,count/(kg\,keV\,day) and
an exposure
of 0.1 and 1\,kg year. For comparison the present limits from the
DAMA
\cite{rita96} and UKDMC \cite{ukdm} NaI experiments are also
shown.}
\label{fig-sapex}
\end{figure}

The present sapphire detectors, with their extremely low energy
thresholds and a
low mass target nucleus with high spin ($Al$),  cover the
low WIMP mass range from 1 GeV to 10 GeV in the sense that
they are presently
the only detectors able to explore this mass range effectively
for non-coherent interactions.
The sensitivity  for WIMPS with spin-dependent interactions, an
expected
 threshold of 0.5 keV, a background of 1\,count/(kg\,keV\,day)  and an
exposure
of 0.1 and 1\,kg year is shown in fig.~\ref{fig-sapex}.
For comparison the present limits from the DAMA \cite{rita96}
and UKDMC \cite{ukdm} NaI experiments are also shown.

Data-taking with the present sapphire ($Al_2O_3$) detectors
(262\,g each)
will continue during 1999.
The goals for this period include the identification and  removal
of noise sources and radioactive contaminants as well as  the
presentation of first results.
In parallel, a  run with a Ge detector is also planned.
Running two target materials in parallel is expected to help
substantially in  understanding
backgrounds and the systematics, as well as preparing the way for
the study of a possible positive signal (see below).

\subsection{Medium and High Mass WIMPs}

\begin{figure}[t]
\vspace*{-2cm}
\begin{center}
\mbox{\epsfig{file=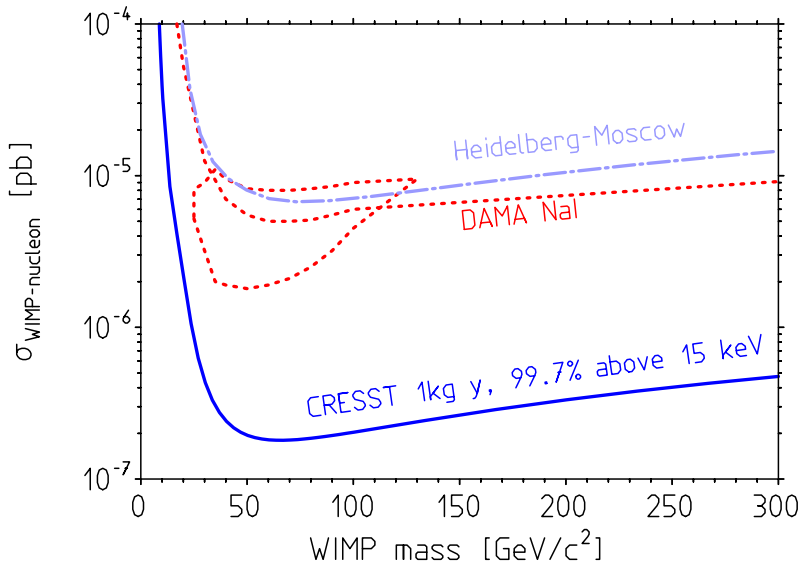,height=13cm,angle=90}}
\end{center}
\caption[]{WIMP-nucleon cross section limits (90\% CL) for scalar (coherent)
interactions
as a function of the WIMP mass, expected for a 1\,kg CaWO$_4$ detector
with a background rejection of 99.7\% above a threshold of 15 keV
detector and  1 year
of measurement time in the CRESST set-up in Gran Sasso.
For comparison the recently updated  limit from the
Heidelberg-Moscow $^{76}$Ge
experiment \cite{hdmnew} and
the DAMA NaI limits \cite{rita96} (with the contour for positive
evidence
\cite{ritapositive}) is also shown.}
\label{fig-caw1kg}
\end{figure}

In the second half of 1999 we intend to start the installation of
the next detector generation
with background suppression using the simultaneous measurement of
scintillation light and phonons. These detectors will have target
nuclei
of large atomic number, such as
tungsten,  making  them particularly sensitive to  WIMPs with
coherent
interactions.
Here the  WIMP cross section profits from a large coherence factor
of the order
$A^2$, ($A=$ number of nucleons).
 Combined with the strong background rejection,
this means these detectors  can be
sensitive to  low WIMP cross sections.
Figure \ref{fig-caw1kg} shows the anticipated sensitivity obtained
with a
CaWO$_{4}$  detector in the present CRESST set-up in Gran Sasso.
The CRESST CaWO$_{4}$ curve is based on a background rate of
1\,count/(kg\,keV\,day),  an intrinsic background rejection of 99.7 \%
above a recoil threshold of 15\,keV and an exposure of 1 kg year.

For comparison the recently updated limits of the Heidelberg-Moscow
$^{76}$Ge-diode  experiment
\cite{hdmnew}, and the DAMA NaI
experiment \cite{rita96} are also shown.  A 60 GeV WIMP with the
cross section claimed in \cite{ritapositive}
would give about 55 counts between
15 and 25 keV in 1\,kg CaWO$_4$ within one year. A background of
1~count/(kg\,keV\,day) suppressed with 99.7\% 
would leave 11 background counts in the same energy range.
A 1\,kg CaWO$_4$ detector with 1 year of measuring time in the
present set-up of CRESST 
should allow a comfortable test of  the recently reported positive
signal.

\subsection {CRESST in the case of a positive signal}

In addition to improving limits on dark matter, it is important to
have
means for  the positive verification of a dark matter signal as
well as for the
elucidation  of its nature.
Once a dark matter signal is suspected, it can be verified by
CRESST through
the following effects.

\begin{itemize}

\item Varying the mass of the target nucleus leads to a definite
shift in the recoil energy spectrum. For example, in the case where
the
WIMP is substantially lighter than the target nucleus, the recoil
{\it momentum}
spectrum has an unchanged shape from nucleus to nucleus.  Hence
there is a simple
rescaling of the recoil energy spectrum.  The observation of
the correct behaviour  will  greatly increase our confidence in
a positive signal. Here  the  significant advantage of the CRESST
technology, that
it can be applied to different target materials, comes
into play.  In this context, the wide variety of materials that may
be used for
simultaneous light and phonon measurement is extremely important.
We have already  measured the relative scintillation efficiencies of
CaWO$_4$, PbWO$_4$, BaF and BGO crystals at low temperatures and
found similarly encouraging results for all materials.

\item Another verification of a dark matter signal is to be
expected  through an annual  modulation of spectral shape and
rate,
which results from the motion of the earth around the sun.
However, a 1\,kg CaWO$_4$ detector is too small to reach a really
significant
statistical accuracy within one year of measurement. Here the
large mass
potential of  the present CRESST installation, about 100 kg,
will play an important role for establishing such an effect in the
future.

\item   Given the detection of a dark matter
particle, an  important task will be  to determine its nature,
e. g. for  SUSY the gaugino and higgsino content, which gives rise
to  different strengths of the
spin-dependent interaction. Significant steps in this
direction  can be taken by using different target materials (see e.g.
fig.~24 in ref.  \cite{pok}).

\item Finally we note calculations \cite{Lkraus}
concerning the possible existence
of a WIMP population orbiting in the solar system rather than in
the galaxy.
These would have a much lower velocity (about 30\,km/sec) as compared to
galactic WIMPs
(270\,km/sec), so that even heavy WIMPs have low momenta.
This underlines the need for low threshold recoil detection and
CRESST is well suited for
such an investigation.

\end{itemize}

\section{Long Term Perspectives}

 The sensitivity reached by a  system  that simply relies passively
on radiopure materials, but  lacks active intrinsic background
suppression, saturates at some point and cannot
 be improved with more mass ($M$) and measuring time ($t$).
On the other hand,
 in a system with a  precisely determined background
suppression factor, the sensitivity
continues to improve  as $\sigma \propto 1/\sqrt{Mt}$
\cite{gaitskell}, as is possible with the CRESST scintillation light
method.

Beginning in the year 2000 we intend to  upgrade the multi-channel
SQUID read out and
systematically increase the detector mass, which can go up to about
100\,kg  before reaching
the full capacity of the present installation. 

\begin{figure}[t]
\vspace*{-1cm}
\begin{center}
\mbox{\epsfig{file=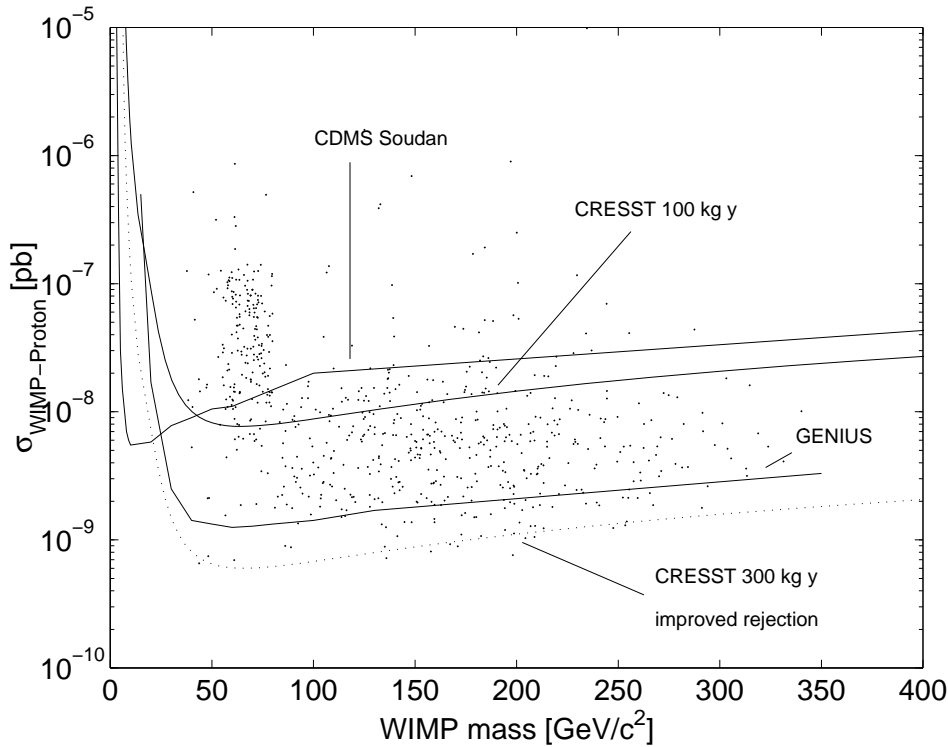,height=10cm,angle=0}}
\end{center}
\caption[]{WIMP-nucleon cross section limits (90\% CL) for scalar (coherent) 
interactions,  as a function of the WIMP mass, expected for a CaWO$_4$ detector
with a
 background of 1~count/(keV kg day) , a background suppression of
99.9\% above
a threshold of 15 keV,  and an exposure of 100\,kg-years in the
CRESST set-up.
With a suppression of 99.99\% above 15 keV, a reduced background of
0.1 counts/(kg keV day), and an increased exposure of 300 kg
years
most of the MSSM parameter space would be covered.
For comparison, the projected sensitivity of CDMS at Soudan
\cite{nam}, and the limits of the proposed GENIUS experiment
\cite{Genius} are also shown. For comparison, all sensitivities are 
scaled to a galactic WIMP density of 0.3 GeV/cm$^3$. 
The dots (scatter plot) represent expectations for WIMP-neutralinos
calculated
in the MSSM framework with non-universal scalar mass unification
\cite{bednyakov}.}
\label{fig-100kg}
\end{figure}

With a 100 kg CaW0$_4$ detector,
the sensitivity  shown in fig.~\ref{fig-100kg}  can be reached in
one year of measuring time.
 If we wish to  cover most of the MSSM parameter space of SUSY with
neutralino dark matter,
the exposure would have to be increased to about 300 kg years, the
background suppression improved to
about 99.99 \% above 15 keV, and the background lowered to 0.1 count/(kg keV
day).  The recent tests in Munich with  CaWO$_4$, which were limited by ambient 
neutrons, 
suggest that a suppression factor of this order should be within reach
underground,  with the neutrons well shielded and employing a muon veto.  

The excellent background suppression of cryodetectors with active
background rejection makes them much less
susceptible to systematic uncertainties than conventional
detectors,
which must rely  heavily on a subtraction of radioactive
backgrounds.
Since this kind of systematic uncertainty cannot be compensated by
an
increase of detector mass, even moderate sized cryogenic detectors
can achieve much better sensitivity than large mass conventional
detectors.
Note that the  excellent levels  shown in fig.~\ref{fig-100kg}   can
be achieved with the
rather moderate assumptions of   background at 1\,count/(kg keV day)
and  0.1 count/(kg keV day).
To a large extent, even higher background levels can be compensated
with increased exposure.  On the other hand, dark matter searches with 
conventional detectors, require a scaling of the presently reached best
background
levels of 0.057 counts/(kg keV day)  \cite{baudis99} by a factor of
2000 to reach the same sensitivity level.

 If WIMPs are not found, at some point  the  neutron flux, which
also gives nuclear recoils, will begin to limit
further improvement.
With careful shielding the  neutron flux in  Gran Sasso should
not limit the sensitivity within the exposures assumed for the
upper CRESST curve in fig.~\ref{fig-100kg}.
With still larger exposures, the  neutron background  may still be
discriminated against large mass WIMPS. This can be done by
comparing  different target materials, which is possible with the
CRESST technology, since different variations with nuclear number
for the  recoil spectra are to be expected with different mass
projectiles.

The phase of the project with large increased total detector mass
will necessitate certain improvements and
innovations in the technology,  particularly involving background
rejection, optimisation of the neutron shielding, and muon vetoing.
 As described above, if a positive dark matter signal exists,
increased mass and
improved background rejection will be important in verifying and
elucidating the effect. 
A large target mass, such as 100\,kg, is of importance to reach the high 
statistics needed to study the annual modulation effect.





\section{Conclusions}

The installation of the large volume, low background, cryogenic facility of 
CRESST at the Gran Sasso Laboratory is completed. 
The highly sensitive CRESST sapphire detectors are up to now the only 
technology available to reasonably explore the low mass WIMP range. 

The new detectors with the simultaneous measurement of phonons and 
scintillation light allow to distinguish the nuclear recoils very effectively  
from the electron recoils caused by background radioactivity. 
For medium and high mass WIMPs this results in one of the highest sensitivities 
possible with today's technology.  

This will allow  a test  of the reported  positive evidence for a WIMP signal by 
the DAMA collaboration in the near future. 
In the long term, the present CRESST set-up permits the installation of
a detector mass up to 100 kg. 
In contrast to other projects, CRESST detectors allow the 
employment of a large variety of target materials.
This offers a powerful tool in establishing a WIMP signal and
in investigating WIMP properties  in the event of a positive signal. 

By its combination of detection technologies CRESST is over the whole WIMP mass 
range one of the best options for direct particle Dark Matter detection.

\end{document}